\documentclass[preprint,12pt]{elsarticle}

\usepackage{graphicx}
\usepackage{amssymb}
\usepackage{tabularx}

\usepackage{lineno}

\journal{ArXiv.org}

\begin{document}

\begin{frontmatter}


\title{Are We Ready for Driver-less Vehicles?\\Security vs. Privacy-\\A Social Perspective}



\author{$Anish Acharya^{1}$}

\address{$^1$University Of California Irvine\\Irvine, CA 92617, United States}

\begin{abstract}
At this moment Automous cars are probably the biggest and most talked about technology in the Robotics Reserach Community. In spite of great technolgical advances over past few years a full fledged autonomous car is still far from reality. This article talks about the existing system and discusses the possibility of a Computer Vision enabled driving being superior than the LiDar based system. A detailed overview of privacy violations that might arise from autonomous driving has been discussed in detail both from a technical as well as legal perspective. It has been proved through evidence and arguments that efficient and accurate estimation and efficient solution of the contraint satisfaction problem adressed in the case of autonomous cars are negatively correlated with the preserving the privacy of the user.  It is a very difficult trade-off since both are very important aspects and has to be taken into account.  The fact that one cannot compromise with the safety issues of the car makes it inevidable to run into serious privacy concerns that might have adverse social and political effects.
\end{abstract}


\end{frontmatter}


\section{Introduction}
\label{sec:introduction}

Over the past few years Autonomous driving has emerged the most challenging task in the Artificial
Intelligence and Robotics Problem. Undoubtedly, it is also the most talked about topic in the automobile
industry these days.
If we look back in the pages of history- autonomous cars existed only in the science fictions. However, the
success of Stanford Robotics group lead by Professor Sebastian Thrun in the 2005 DARPA grand Urban
Challenge ~\cite{thrun2006stanley} first proved that it is not too far that autonomous cars can be a reality. Ever since 2005 the tech
giant “Google” started working on this and recently announced the first working, scalable driver-less car.
Now, automobile giants are spending billions of dollar on autonomous driving technology.
As every technological innovation is driven by Social needs the motivation behind developing driver-less
cars were also driven by social needs like Road Safety, better traffic monitoring, driver safety and many
other as mentioned by the Sebastian Thrun of Google X .However, as it became a reality many questions are being posed by the robotics community as well as the 
informatics and social community. 
\\In spite of the huge hype surrounding Autonomous car,  we believe that we are at-least a decade behind having a fully functional autonomous car in terms of technology. The current system can only drive at a max speed of 25 miles/hr. Many loop holes exist including its performance in the urban jungle. Its performance in Freeway is also questionable.
\\
Almost all existing autonomous vehicles rely on active sensing (eg. lidar) to construct point cloud representation of the surrounding scene.
However, due to hardware limitations existing LiDar framework cannot spin at afrequency more than 10-15 Hz. Due to this reason the point cloud representation of the surrounding obtained from the LiDar information is often very sparse and more so in case of speeding or in case of complex/cluttered surrounding. 
With the recent successes of vision based algorithms in accurately predicting the scene, there is a strong indication that passive computer vision has the potential to provide equally rich geometrical representation of surrounding at a lower cost.
\\Rather it can be inferred that a combination of both Physical sensor based framework with vision enabled framework would improve the efficiency of the car by quite a few notch. Though, it is still an open question but from an information theoretic point of view more data from the surrounding would lead to a better inference and decision making by the car.
\\However, there is one issue that has to be considered. It is the privacy of the users that also needs to be taken into account. Privacy is one of the basic rights and not a privilage. ~\cite{warren1890right} Thus, it is the primary responsibility to ensure that autonomous cars do not violate privacy of an individual. There are many questions that are arising on this issue because the autonomous car makes use of several technologies that might lead to violation of privacy rights. 
\\Privacy and surveillance issues may arise from the use of GPS and the interconnection between different cars of same kind (C2C) and also between  car and device/ with Law enforcement (C2D) etc. 
\\ This article moves around these issues. The central argument being to evaluate how there exists a trade-off between improving the accuracy/prediction of the autonomous car and privacy preservation.
\\ We discuss how Vision enabled systems, in spite of being technologically improve the performance of the whole pipeline but it might turn out to be a disaster in terms of being privacy preserving.  
\\

The rest of the paper is organised as follows:
Section 2 starts by describing the motivation behind autonomous vehicle research and is supported by significant amount of literature review and statistical evidences. We then give a brief overview of how autonomous cars evolved over past few years and give a very brief semi-technical overview of the basic working principle of current system. Next, a semi-technical discussion on how vision enabled system might perform better than the current LiDar based system has been provided. Section 3 starts by discussing the Privacy concerns that might be associated with Driver-less vehicles.A detailed analysis has been done both from a technical and from a legal perspective on the potential privacy threats in the current LiDar based sytem and the additional privacy concerns coming from Vision enabled autonomous cars. The section ends with a discussion on how the trade-off between privacy and performance makes it difficult for autonmous cars to be developed in its full glory. The last section wraps up the argument posing some open questions to be adressed in future.   

\section{Are we ready for Autonomous Vehicles?}
\label{sec:section_1}
Over the past few decades road fatalities have been a major cause of concern for USA and around the globe. Fig.1 shows the statistics of deaths in United States over past few years as published by The 2012 Statistical Abstract published by United states Cencus Bureau.
\begin{figure*}[ht]
\centering
\includegraphics[width=12cm,height=6cm]{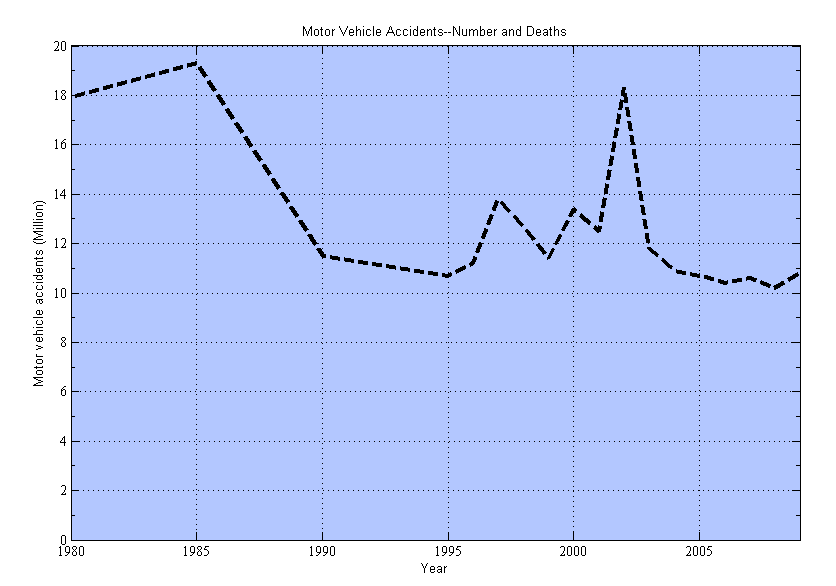}
\caption{Motor Vehicle Accidents--Number and Deaths}
\end{figure*}
As indicated by this report it is an alarming situation for the whole nation. It can be readily inferred from the data that in spite of multifold improvement in traffic control over the period 1980-2010 road fatalities have merely been in control. Further investigation into the reports released indicate that most of these accidents are due to human error rather than instrumental failures. This has been the primary motivation behind the extensive research on Autonomous vehicles over the past few years. The hope is that the more we move towards automation the less will be the probablity of human error leading to a decrease in road fatalities and should lead to new levels of improved traffic operation including comfortable and safe driving experience. The central motivation behind the development of Autonomous Cars have been to monitor traffic in a regularized manner and thus to reduce possiblity of accidents and  ensuring improved road safety. It is already becoming more and more popular and likely to be seen more and more on the roads of US, with some states like Michigan,California,Nevada and Florida having already introduced legislation authorizing Autonomous cars. With great prospects, comes greater responsibilities. Since, people will be the intended users and consumers of Autonomous vehicles,safety and privacy becomes two major concerns and most important aspects to be taken into consideration before making its way to the roads.Vanderbilt et.al. suggests this to be "the most slippery territory for autonomous vehicles" among all other social and cultural issues~\cite{vanderbilt2012let}. In this section we carefully investigate this two issues and try to understand where do we stand in terms of moving towards a completely autonomous driving experience. 

\subsection{\bf Towards a safer Driving Experience}
The primary reason behind trying to move towards a more and more autonmous driving infrastructure is to reduce road fatalities. Over the past few years there has been a lot of research towards the goal of fully autonomous car in an incremental way~\cite{milanes2010clavileno}.The idea has been to start from automating simple driving aids and scale upto completely automatic controls. Many high range cars are already equipped with tools like Cruise Control(CC) that allows the driver to cruise at a fixed speed and its advanced version Adaptive Cruise control (ACC) in which mode the car automatically adjusts its speed in accordance with the other cars in the road. Many other aids like Collision avoidance system,Safety alert seat,Lane departure warning,Rear vision camera with dynamic guidelines,Back-up collision intervention etc are already in action and are very popular. It is not hard to see that these if properly employed would lead to improved safety or atleast would be able to keep the driver more informative about the surrounding and may also warn drivers about possible mistakes they might be making. With subsequent advancement in research the cars have become more and more safe,efficient and robust as expected. As mentioned by Black et.al. ~\cite{black1996sustainable} National Highway Traffic Safety Administration (NHTSA) published an official classification rule to classify the level of automation in different stages of development towards the fully atonomous car. The main overall categories being-
\begin{itemize}
\item {\bf Level 0:} The era of completely manual driving falls under this category. There is no doubt that we have left this stage long back.
\item {\bf Level 1:} Individual Control automated. This refers to the early stage of Autonomous Car research when we saw the emergence of automated driving tools one by one.
\item {\bf Level 2:} This is the stage where there are atleast two or more automative controls employed in the car and work in unision. 
\item {\bf Level 3:} In this generation of autonomous cars the driver would be able to fully cede control under certain circumstances and the car would be able to sense the conditions where the driver should take over the control and allows the driver with sufficient time to take over.
\item {\bf Level 4:} This is the long awaited stage- where the car can sense and take decisions on its own in a fully automatic manner and does not require the driver to intervene at any time. It should be fully aware of the surrounding and "smart" enough to act during emergencies.  

\end{itemize}

Due to the enormous amount of research by the Robotics Community over the past few decades we have passed the first three stages successfully and the current research focus is on the Fourth generation of autonomous cars. Recently the tech giant Google came up with the first commercial fully autonomous car~\cite{markoff2010google} (Fig 2).
Though, it is a great achievment in the ongoing robotics research for autonomous car development; it has arouse many eyebrowse and there are still many unanswered questions. 
We strongly believe that it is just the first step towards the goal of fully autonomous cars and we are atleast a few decades behind when we should be able to see a fully autonomous car in its full glory on the roads. In order to understand why a fully autonomous car is still far from commercial use we have to understand the technical details of the current system.     
\begin{figure*}[ht]
\centering
\includegraphics[width=12cm,height=6cm]{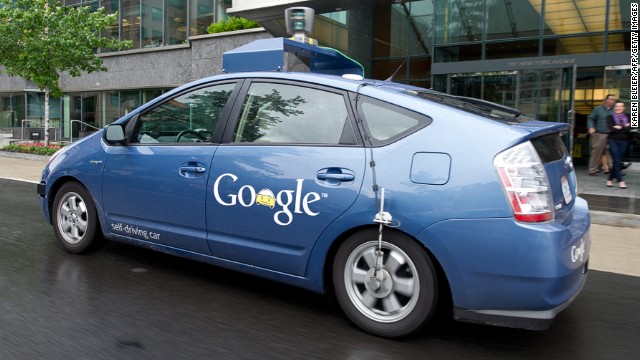}
\caption{Google's autonomous car in action}
\end{figure*}
On a closer look, it can be observed that there is a LiDar mounted on top of the car. It is the most important active sensing used in the car. There are many variants of LiDar in the market, however Velodyne HDL-64E being the most popular one used in driver-less vehicles ~\cite{halterman2010velodyne}. It consists of 128 lasers split into two blocks-upper and lower block. (fig.3)~\cite{schwarz2010mapping}
\begin{figure*}[ht]
\centering
\includegraphics[width=6cm,height=6cm]{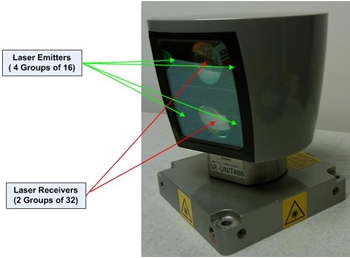}
\includegraphics[width=6cm,height=6cm]{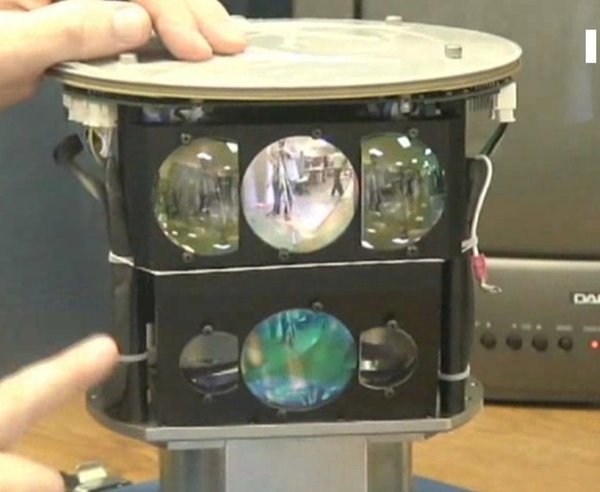}
\caption{Velodyne LiDar}
\end{figure*}
The velodyne typically rotates at a frequency of 10-15 Hz. It uses SONAR principle to get a depth estimation of the surrounding. As the laser hits the objects in the surrounding and comes back to the recievers of the LiDar it basically uses the time of flight information to predict the depth map of the surrounding. Though apparantly it looks like fetching a lot of information about the surrounding,however a closer look immediately reveals that the point cloud obtained from the LiDar spin is extremely sparse and it cannot be considered to be extremely informative. It can be immediatly extrapolated that being less informative about the surrounding it is extremely hard for the car to make accurate predictions about the surrounding. The problem is more accurate in cases where - firstly the car is at a higher speed and secondly when there are larger number of objects especially moving objects. If considered practically this is something that might be very common in free-ways. Thsi is the reason why the Google car hasn't been able to drive safely at a speed more than 25 miles/hr~\cite{milanes2010clavileno}. Another potential area to consider is over-crowded street and unordered traffic and pedestrian. Since, the current system depends on active sensing it is very difficult for it to perform inference based on only point cloud. As an argument it may be coined that the car is generally equipped with GPS and a pre-learned map of the area available from satelite images. However, one point to be noted here is that the situations are not so ideal everytime. Pedestrian might include children, also there might be someone who is drunk or someone aged. Undoubtedly, under these different conditions it is almost impossible for the car to adapt and act accordingly with the current state of artificial intelligence research. 

\subsection{\bf My Car has Eyes}
It can readily be observed that in order to make better decision the car has to more informed about its surrounding. From an information theoretic point of view it can be concluded that Computer Vision holds the potential to replace and perform better than active sensor based systems. Off late there has been an emerging trend among the robotics community to switch to Vision based navigation framework instead of relying entirely on active sensing like LiDar. With the recent
successes of vision based algorithms in accurately predicting the scene, there is
a strong indication that passive computer vision has the potential to provide
equally rich geometrical representation of surrounding at a lower cost.Also, in a computer vision based framework object tracking,labelling becomes far more trivial as compared to the point cloud representation where individual labelling is difficult and inaccurate. 

In Vision based framework the idea is to reconstruct the surrounding
scene as accurately as possible. This itself is a big research problem and we
are still far from predicting the 3D scene exactly. However, recent success of
slanted plane based models in very challenging KITTI benchmark ~\cite{geiger2013vision} proves the
superiority of Slanted plane models to reconstruct the 3D scene of the surrounding.~\cite{yamaguchi2013robust},~\cite{yamaguchi2012continuous}.
These slanted plane methods are based on the basic assumption
that the 3D scene is piece-wise planar and the motion is piece-wise rigid. Where
these earlier works used more expensive inference algorithm such as particle belief
propagation, a more recent work by Yamaguchi et. al. ~\cite{yamaguchi2014efficient} proposed a much
faster algorithm based on stereo and video pairs to jointly estimate depth,image
segmentation as well as boundary labels. It uses block-coordinate descent on
the total energy involving the segmentation. The whole process takes around 3
seconds to converge to global optimum- thus giving rise to a practical slanted
plane model of the surrounding 3D scene.
\\This algorithm is capable of taking disparity image as input and constructs an image segmentation at super-pixel level and fits a slanted plane for each segment. 
The slanted plane thus obtained is parametrized by (a,b,-1,d) given as eq. (1) 
\begin{equation}
ax+by-z+d=0
\end{equation}
conversion from slanted plane to 3D plane is based on the standard pinhole camera model.There is a transformation between the 3D world and 2D image coordinates together with disparity, based on an projection matrix. Hence, disparity planes can be converted to 3D scene planes using the standard camera-to-world transform. We use the following relation for this conversion between image co-ordinates x,y and spac co-ordinates X,Y,Z. 
\begin{equation}
x=f_x\frac{X}{Z}+o_x
\end{equation}
\begin{equation}
y=f_y\frac{Y}{Z}+o_y
\end{equation}
\begin{equation}
z=b\frac{f_x}{Z}
\end{equation}
Where $f_x,f_y,o_x,o_y$ are standard camera intrinsic and z is the disparity. 
It can be shown that the resulting 3D plane equation is given as (5)
\begin{equation}
af_xX+bf_yY+(ao_x+bo_y+d)z-bf_x=0
\end{equation}
The set of segments and parameters are computed based on an energy-minimization approach using stereo map and in the meantime accounting for other factors such as segment shape, plane smoothness etc. Fig.5 provides two example of slanted plane algorithm to infer depth. 
\begin{figure*}[ht]
\centering
\includegraphics[width=15cm,height=3cm]{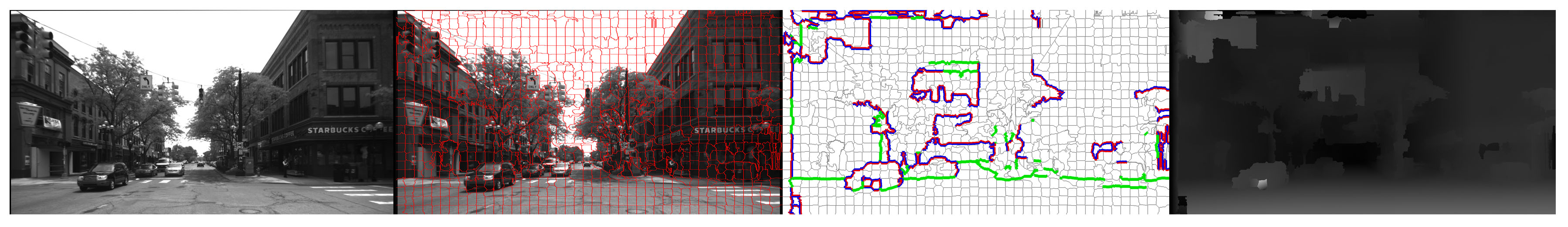}
\caption{. The input image and the corresponding smoothed version}
\end{figure*}
\begin{figure*}[ht]
\centering
\includegraphics[width=15cm,height=3cm]{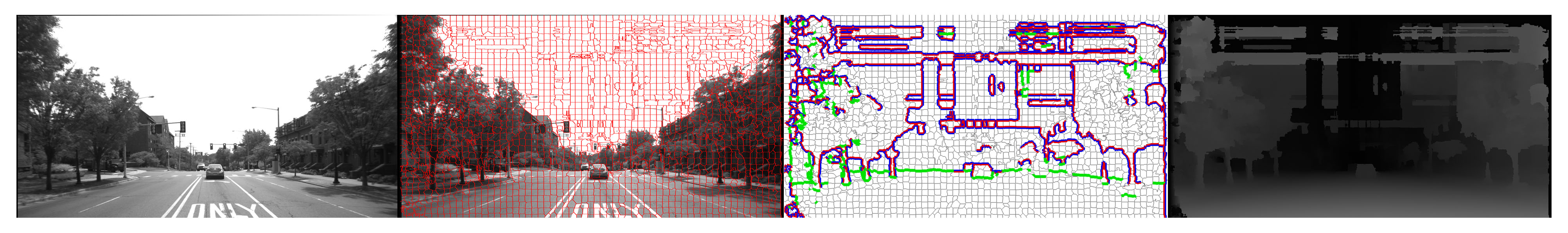}
\caption{. The input image and the corresponding smoothed version}
\end{figure*}

The KITTI benchmark suite is an important step towards the research involving Vision enabled car. It provides a platform to compare the accuracy of depth estimates of a vision based algorithm with respect to the ground truth point cloud obtained from the physical sensor measurement. 
Table 1 reveals the comparative performance of different Vision based framework on the KITTI benchmark~\cite{geiger2013vision}.
\\
Most of the research in the autonomous driving community is around the application of computer vision based algorithm to enable rich,more informative geometric representation of the surrounding for better inference of the car. Even the Asian Conference in Computer Vision (ACCV 2014) held in Singapore had a separate workshop and tutorial session only on the application of Vision framework in Autonomous Car.

\section{Security Vs Privacy}
\label{sec:section2}
As with other emerging disruptive technologies,like drones and wearables, it is essential that issues relating to user privacy and data security are properly addressed prior to the technologies being generally deployed. As mentioned by Glancy et.al."If people were not involved with autonomous vehicles,
privacy would not be an issue"~\cite{glancy2012privacy}.Autonomous car in spite of its great prospects in terms of safe,comfortable driving experience has been  criticized by many accross the globe for the potential privacy threat to the users of autonomous cars and the people around it. As discussed in the previous section the main technological concern has been to make the car as safe and make the predictions as accurate as possible.
\\since, the LiDar based system is still the state-of-the-art, it is often not enough from a safety perspective. Though, Autonomous cars are not controlled by any central system, from a safety perspective it is equipped with navigation system. As mentioned by  Lee et.al. in a recent article appearing in The Washington Post-
"Fights over privacy in the self-driving future will focus on many of the same issues that are currently being debated with regard to cellphones".\footnote{Self-driving cars are a privacy nightmare. And it’s totally worth it.By Timothy B. Lee May 21, 2013 © 1996-2014 The Washington Post} 
\\Though the fact that it is not controlled from a central server might indicate that it is not being tracked but considering the fact that while one car drives without driver, it needs to be connected through cellular network to navigate. This will allow it to pick/drop-off passengers at the right place which is the concept of autonomous public transport. The whole system is a big source of private data and at any point it is trivial to pull up the exact location and efficiently track anyone giving rise to a huge privacy concern.
\\There are two standard protocols available for communicating between autonomous vehicles. namely V2V and V2D. 
In vehicle-to-vehicle (V2V) protocol the vehicle receives and shares its internal data with other similarly connected, data-sharing vehicles\footnote{In the USDOT “Connected Vehicle Program,” the Core (or Core System),
is such an enterprise network for communications among vehicles and between
vehicles and other elements participating in the Core, such as traffic
management, navigation applications and many other potential users.
Connected Vehicle Core System Baseline Documentation, U.S. DEP’T OF TRANSP.:
RESEARCH and INNOVATIVE TECH. ADMIN.}. 
Wheras the Vehicle to Device (V2D) or Vehicle to Infrastructure(V2I) where the vehicle shares its internal data ((speed, velocity, heading, etc.) with Law Enforcement agencies, traffic management centers etc.  Independent of which protocol being used all pose serious threat to privacy. "Different types and degrees of privacy
protection will be needed depending on the types of technologies represented by the two models of autonomous
vehicles discussed here, or whatever combination of
technologies is eventually built into future autonomous
vehicles"~\cite{glancy2012privacy}.   
\\In summary there are three types of major privacy concerns personal autonomy, personal
information, and surveillance.
\subsection{\bf personal autonomy}
From a Law perspective, personal autonomy comes with many types of privacy rights of an individual. However surprising it may sound in an era of digital personal information and the Internet, autonomous cars might cause greater violation of personal autonomy. 
\\Firstly, whether or not to choose an autonomous vehicle is an individual's primary right. It is concerned with "individual control and
self-determination people’s abilities to make independent
choices about themselves"\footnote{Autonomia, n., OXFORD ENGLISH DICTIONARY (2d ed. 1989) (Oxford
Univ. Press).} \footnote{IMMANUEL KANT, THE CRITIQUE OF PRACTICAL REASON (L. W. Beck,
trans., Macmillan Publishing Co. 3d ed. 1993).} \footnote{THOMAS AQUINAS, BASIC WRITINGS OF ST. THOMAS AQUINAS, (A. C.
Pegis, ed., Hackett Publishing Co. 1997).}
However, if autonomous cars are to operate properly and in order to ensure regulated traffic all the cars in the road has to be equally smart and communicating with each other. Having said that, it would be insane if half of the cars are autonomous/driverless and half have human drivers, since that would be chaotic due to the difference in decision making of human and robot. 
\\However the most biggest concern is why should the owner be responsible for the accidents caused by autonomous cars. This is a big loop-hole since if one has to be blamed it has to be the owner because it is impractical to blame the robot. ~\cite{gurney2013sue}. Also,the insurance rate is determined by the driving history of the driver whereas in case of autonomous cars the owner is forced to take the liability of the accidents caused by the car and also forced to pay the insurance rate evaluated based on the efficiency of the car since "Applying autonomy to a non-human vehicle is shamelessly anthropomorphic"~\cite{gurney2013sue}. 
\\Another almost unavoidable situation is to maintain the anonymity of the person travelling which is an important right of people. However, in case of connected autonomous vehicle anonymity might raise security concerns about "being able to trace
misbehaving technology, or to find antisocial activity or to
prosecute individuals responsible for unlawful network
activities." But it is against the recent judgment by US supreme court that recognizes the right to choose anonymous personal mobility. \footnote{United States v. Jones, 132 S. Ct. 945 (2012).}
\subsection{\bf Surveillance Privacy Interests}
"Surveillance privacy interests respond to people’s
aversion to being constantly watched, tracked or monitored as
they travel from place to place. At the same time,
surveillance privacy interests also reflect political and
philosophical opposition to pervasive scrutiny of everyone
who travels, particularly if that scrutiny is controlled by
government"\footnote{United States v. Jones, 132 S. Ct. 945, 956 (2012) (Sotomayor, J.,
concurring).}. The potential use of autonomous vehicles for surveillance may be much bigger cause of privacy concern than peronal or informational privacy. It holds the potential to threaten social and political well-being of the society. As Supreme
Court Justice Sonia Sotomayor noted "Awareness that the
Government may be watching chills associational and
expressive freedoms. And the Government’s unrestrained
power to assemble data that reveal private aspects of identity
is susceptible to abuse.” She also pointed out making
available at a relatively low cost such a substantial quantum
of intimate [GPS location] information about any person
whom the Government, in its unfettered discretion, chooses to
track," may "alter the relationship between citizen and
government in a way that is inimical to democratic
society."\footnote{Id. (citing United States v. Cuevas-Perez, 640 F.3d 272, 285 (7th Cir.
2011) (Flaum, J., concurring)).} 
There are two types of surveillance- covert and overt. Though surveillance often means covert collection of information but in case of autonomous cars it is mostly overt surveillance similar to the red light cameras. The main purpose is to affect the behaviour of the subject. It can be the internal data being reported to the Law enforcement or instances of behaviour like smoking,drinking can be recorded. 
The purpose of overt surveillance is "interfere with
individual autonomy through the power of scrutiny"\footnote{See MICHAEL FOUCAULT, DISCIPLINE AND PUNISH: THE BIRTH OF THE PRISON (1979).}

\subsection{\bf Personal Information Privacy}
The more informative and accurate we wish the car to be the more we violate the privacy concerns. It can be readily observed that in order to improve the decision making of the autonomous car the car needs to be more informative about its surounding which is directly correlated to the violation of privacy. For this reason the cars become a great source of data. Even if the owner of the car is compromises with his own personal autonomy the problem is that it is not only confined to the owner or the users of the car, the autonomous car gathers large amount of data of people around it.
\\ It is very common to use private, sensitive, personal data in many research discipline. Consider a research on autism, or a research on the social behaviour of people, or the social experiments that goes on via social networks. It is all over the internet and the IRB is happy as long as there are no tags to the data. It is very common in clinical studies to conduct experiments, surveys on people suffering from a desease or people receiving a particular treatment. However, while recording the data all tags or labels associated with the data are not recorded or removed to maintain anonymity. However, there may not be such an easy escape in case of autonomous vehicles. 
\\This becomes a greater cause of concern when we look into the recent trend in Robotics research to move towards a Vision Enabled autonomous cars discussed in Section 2, due to its better accuracy in prediction and thus leading to more accurate decision making. At first glance it might not be so obvious to see the difference between the LiDar based and the Vision based approach in terms of privacy concerns. However, a closer look would reveal that Vision enabled cars pose a much greater threat to people's privacy. The vision enabled system relies on images taken from multiple cameras. From the principle of multiview geometry it is possible to re-construct the entire 3D surrounding based on the relative position of the cameras installed on the car and the corresponding 2D images. All the Computer Vision based framework uses this fundamental aslgorithm in its heart. Which implies that any Vision enabled autonomous vehicle will take multiple images of everything within the Field of Vision(FOV) of the camera from different angles at a frequency similar to that of the LiDar. 
\\

In case of purely point cloud based representation of the world around the car, it is not that it is completely anonymous because anyways the point cloud is associated with the GPS information the car has. However, there is a slim chance while storing the log if it the point cloud data is stored without a time-stamp. However, in case of Vision enabled framework it is very hard to annonymize data. The images capture everything around the car which might be the cause of privacy violance of not only the owner of the car but the people around it. Some examples include- number plates of cars, capturing the activity of the pedestrians, name of shops on the side of the road, activity of people standing beside the road, even activity of people inside the shops etc. These are huge violation of privacy interests and should be taken care of before moving towards a vision enabled autonomous driving. 
\\

It can be inferred from the discussion so far that in spite of much hype of Google's autonomous car, from a practical and purely technical perspective it is still far from replacing the traditional human-driven cars. There are many things unanswered, for example the car is still not capable of driving itself succesfully on a over crowded street, the max speed achieved is still limited to 25-30 miles/hour. We discussed in Section 2 that this can mainly be attributed to the limitation of physical range sensors. From a hardware design perspective it is very difficult to come up with a LiDar that rotates at a very high speed without loss of accuracy which directly relates to the sparse nature of the generated point cloud of the surrounding. This limitation can somehow be better adressed by Computer Vision , where vision enabled cars obtain a richer geometric representation of its surrounding based on the 3D reconstruction of surrounding from multiple camera images taken simultaneously from different angles using various "depth from motion" algorithms. However, we argue that this leads to a greater potential threat to privacy and it is very difficult to ananymise the huge amount of structured data obtained from the car. This extends to greater political and social well-fare of people and there is also an angle of surveillance/ monitoring here. In one way, it might be useful from the perspective of the Law enforcement as they can monitor each and everyone all the times but that causes a major intrusion to people's privacy. 
\\We can easily extrapolate that neither LiDar based system, nor the vision based system can fully solve the problem of Autonomous driving. It would be a combination of physical range sensors like LiDar, data coming from multiple camera mounted in the car, GPS, navigation tools, wireless connection, interconnectivity of different cars and connectivity between  Car and Devices etc would have to be employed together in order to reach the desired fully autonomous car as depicted in the sci-fi movies.
\\However, it should also be kept in mind that each of these
additions gives rise to more and more data which in turn
gives rise to different kinds of and different levels of privacy violation. 
This leads to a trade-off between improved performance vs. less privacy violation. And, this might be one of major challenges in the research and development of autonomous driving and it is very less likely to see fully- autonomous cars in action in recent future. 



\section{Conclusion}
In summary it can be inferred that it is a very difficult trade-off between security and privacy. In order to make a safer, more robust, scalable autonomous car information is required. The more information the car gets about its surrounding the more aware it is with respect to what are the constraints it has to take care of which will automatically improve the inference and decision making of the car. This is directly in line with the rule of statistical learning. At the end it is basically a statistical inference problem in a semi-supervised learning framework where the car is basically trained with respect to training data and infers based on the learned constrained satisfaction problem. Along the lines of any constraint satisfaction problem it is helpless in absense of sufficient data. Thus, in order to make the car more efficient and reliable it has to be fed with more information of the surrounding , which in turn leads to a greater threat to user privacy. 
\\As of now we stand at a point in the research and development of Level 4 of Autonomous cars, where it is at a very critical balance between better performance and privacy violation. And it looks like that the biggest challenge in the coming decade is to come up with a learning strategy that uses data in an anonymous manner.
\\ The big question for future researches in the robotics field would be to find a "privacy preserving constraint satisfaction algorithm" that would lead to a safe , efficient, robust yet privacy preserving fully Autonomous vehicle. 




\section{References}
\bibliographystyle{model1-num-names}
\bibliography{sample.bib}

\begin{thebibliography}{14}
\expandafter\ifx\csname natexlab\endcsname\relax\def\natexlab#1{#1}\fi
\providecommand{\bibinfo}[2]{#2}
\ifx\xfnm\relax \def\xfnm[#1]{\unskip,\space#1}\fi
\bibitem[{Thrun et~al.(2006)Thrun, Montemerlo, Dahlkamp, Stavens, Aron, Diebel,
  Fong, Gale, Halpenny, Hoffmann et~al.}]{thrun2006stanley}
\bibinfo{author}{S.~Thrun}, \bibinfo{author}{M.~Montemerlo},
  \bibinfo{author}{H.~Dahlkamp}, \bibinfo{author}{D.~Stavens},
  \bibinfo{author}{A.~Aron}, \bibinfo{author}{J.~Diebel},
  \bibinfo{author}{P.~Fong}, \bibinfo{author}{J.~Gale},
  \bibinfo{author}{M.~Halpenny}, \bibinfo{author}{G.~Hoffmann}, et~al.,
\newblock \bibinfo{title}{Stanley: The robot that won the darpa grand
  challenge},
\newblock \bibinfo{journal}{Journal of field Robotics} \bibinfo{volume}{23}
  (\bibinfo{year}{2006}) \bibinfo{pages}{661--692}.
\bibitem[{Warren and Brandeis(1890)}]{warren1890right}
\bibinfo{author}{S.~D. Warren}, \bibinfo{author}{L.~D. Brandeis},
\newblock \bibinfo{title}{The right to privacy},
\newblock \bibinfo{journal}{Harvard law review}  (\bibinfo{year}{1890})
  \bibinfo{pages}{193--220}.
\bibitem[{Vanderbilt(2012)}]{vanderbilt2012let}
\bibinfo{author}{T.~Vanderbilt},
\newblock \bibinfo{title}{Let the robot drive: The autonomous car of the future
  is here},
\newblock \bibinfo{journal}{Retrieved October} \bibinfo{volume}{9}
  (\bibinfo{year}{2012}) \bibinfo{pages}{2013}.
\bibitem[{Milan{\'e}s et~al.(2010)Milan{\'e}s, Llorca, Vinagre, Gonz{\'a}lez,
  and Sotelo}]{milanes2010clavileno}
\bibinfo{author}{V.~Milan{\'e}s}, \bibinfo{author}{D.~F. Llorca},
  \bibinfo{author}{B.~M. Vinagre}, \bibinfo{author}{C.~Gonz{\'a}lez},
  \bibinfo{author}{M.~A. Sotelo},
\newblock \bibinfo{title}{Clavile{\~n}o: Evolution of an autonomous car},
\newblock in: \bibinfo{booktitle}{Intelligent Transportation Systems (ITSC),
  2010 13th International IEEE Conference on}, \bibinfo{organization}{IEEE},
  pp. \bibinfo{pages}{1129--1134}.
\bibitem[{Black(1996)}]{black1996sustainable}
\bibinfo{author}{W.~R. Black},
\newblock \bibinfo{title}{Sustainable transportation: a us perspective},
\newblock \bibinfo{journal}{Journal of Transport Geography} \bibinfo{volume}{4}
  (\bibinfo{year}{1996}) \bibinfo{pages}{151--159}.
\bibitem[{Markoff(2010)}]{markoff2010google}
\bibinfo{author}{J.~Markoff},
\newblock \bibinfo{title}{Google cars drive themselves, in traffic},
\newblock \bibinfo{journal}{The New York Times} \bibinfo{volume}{10}
  (\bibinfo{year}{2010}) \bibinfo{pages}{A1}.
\bibitem[{Halterman and Bruch(2010)}]{halterman2010velodyne}
\bibinfo{author}{R.~Halterman}, \bibinfo{author}{M.~Bruch},
\newblock \bibinfo{title}{Velodyne hdl-64e lidar for unmanned surface vehicle
  obstacle detection},
\newblock in: \bibinfo{booktitle}{SPIE Defense, Security, and Sensing},
  \bibinfo{organization}{International Society for Optics and Photonics}, pp.
  \bibinfo{pages}{76920D--76920D}.
\bibitem[{Schwarz(2010)}]{schwarz2010mapping}
\bibinfo{author}{B.~Schwarz},
\newblock \bibinfo{title}{Mapping the world in 3d},
\newblock \bibinfo{journal}{Nat. Photonics} \bibinfo{volume}{4}
  (\bibinfo{year}{2010}) \bibinfo{pages}{429--430}.
\bibitem[{Geiger et~al.(2013)Geiger, Lenz, Stiller, and
  Urtasun}]{geiger2013vision}
\bibinfo{author}{A.~Geiger}, \bibinfo{author}{P.~Lenz},
  \bibinfo{author}{C.~Stiller}, \bibinfo{author}{R.~Urtasun},
\newblock \bibinfo{title}{Vision meets robotics: The kitti dataset},
\newblock \bibinfo{journal}{The International Journal of Robotics Research}
  (\bibinfo{year}{2013}) \bibinfo{pages}{0278364913491297}.
\bibitem[{Yamaguchi et~al.(2013)Yamaguchi, McAllester, and
  Urtasun}]{yamaguchi2013robust}
\bibinfo{author}{K.~Yamaguchi}, \bibinfo{author}{D.~McAllester},
  \bibinfo{author}{R.~Urtasun},
\newblock \bibinfo{title}{Robust monocular epipolar flow estimation},
\newblock in: \bibinfo{booktitle}{Computer Vision and Pattern Recognition
  (CVPR), 2013 IEEE Conference on}, \bibinfo{organization}{IEEE}, pp.
  \bibinfo{pages}{1862--1869}.
\bibitem[{Yamaguchi et~al.(2012)Yamaguchi, Hazan, McAllester, and
  Urtasun}]{yamaguchi2012continuous}
\bibinfo{author}{K.~Yamaguchi}, \bibinfo{author}{T.~Hazan},
  \bibinfo{author}{D.~McAllester}, \bibinfo{author}{R.~Urtasun},
\newblock \bibinfo{title}{Continuous markov random fields for robust stereo
  estimation},
\newblock in: \bibinfo{booktitle}{Computer Vision--ECCV 2012},
  \bibinfo{publisher}{Springer}, \bibinfo{year}{2012}, pp.
  \bibinfo{pages}{45--58}.
\bibitem[{Yamaguchi et~al.(2014)Yamaguchi, McAllester, and
  Urtasun}]{yamaguchi2014efficient}
\bibinfo{author}{K.~Yamaguchi}, \bibinfo{author}{D.~McAllester},
  \bibinfo{author}{R.~Urtasun},
\newblock \bibinfo{title}{Efficient joint segmentation, occlusion labeling,
  stereo and flow estimation},
\newblock in: \bibinfo{booktitle}{Computer Vision--ECCV 2014},
  \bibinfo{publisher}{Springer}, \bibinfo{year}{2014}, pp.
  \bibinfo{pages}{756--771}.
\bibitem[{Glancy(2012)}]{glancy2012privacy}
\bibinfo{author}{D.~J. Glancy},
\newblock \bibinfo{title}{Privacy in autonomous vehicles},
\newblock \bibinfo{journal}{Santa Clara L. Rev.} \bibinfo{volume}{52}
  (\bibinfo{year}{2012}) \bibinfo{pages}{1171}.
\bibitem[{Gurney(2013)}]{gurney2013sue}
\bibinfo{author}{J.~K. Gurney},
\newblock \bibinfo{title}{Sue my car not me: Products liability and accidents
  involving autonomous vehicles},
\newblock \bibinfo{journal}{U. Ill. JL Tech. \& Pol'y} \bibinfo{volume}{13}
  (\bibinfo{year}{2013}) \bibinfo{pages}{247--423}.

\end{thebibliography}








\end{document}